\newcommand{\CLASSINPUTtoptextmargin}{1.91cm}
\newcommand{\CLASSINPUTbottomtextmargin}{1.91cm}
\documentclass[conference, a4paper]{IEEEtran}
\IEEEoverridecommandlockouts

\usepackage{cite, url}

\usepackage{color,soul}
\usepackage{colortbl}
\usepackage[T1]{fontenc}

\usepackage{tikz, forest}
\usepackage{etoolbox}

\ifCLASSINFOpdf
\usepackage{graphicx}
\usepackage{tabularx}
\usepackage{amssymb}
\usepackage{float}
\let\labelindent\relax
\usepackage{enumitem}
\graphicspath{{./pdf/}{./jpeg/}}
\else
\fi

\usepackage{algorithm}

\usepackage[noend]{algpseudocode}

\makeatletter

\newsavebox{\measure@tikzpicture}
\NewEnviron{scaletikzpicturetowidth}[1]{%
  \def\tikz@width{#1}%
  \def\tikzscale{1}\begin{lrbox}{\measure@tikzpicture}%
  \BODY
  \end{lrbox}%
  \pgfmathparse{#1/\wd\measure@tikzpicture}%
  \edef\tikzscale{\pgfmathresult}%
  \BODY
}
\renewcommand{\ALG@beginalgorithmic}{\footnotesize}
\def\BState{\State\hskip-\ALG@thistlm}
\makeatother

\usetikzlibrary{arrows.meta}
\usetikzlibrary{backgrounds}
\usetikzlibrary{matrix}

\ifCLASSOPTIONcompsoc
  \usepackage[caption=false,font=normalsize,labelfont=sf,textfont=sf]{subfig}
\else
  \usepackage[caption=false,font=footnotesize]{subfig}
\fi

\usepackage{array}
\usepackage{pbox}

\usepackage[cmex10]{amsmath}
\usetikzlibrary{calc}
\usetikzlibrary{positioning}
\usetikzlibrary{decorations.text}
\usetikzlibrary{arrows.meta}
\usepackage{multirow}

\usepackage[absolute,showboxes]{textpos}
 
\TPMargin{5pt}

\begin{document}
%
\title{BFT Protocols for Heterogeneous Resource Allocations in Distributed SDN Control Plane}

\vspace{3cm}
\author{
\IEEEauthorblockN{Ermin Sakic\IEEEauthorrefmark{1}\IEEEauthorrefmark{2}, Wolfgang Kellerer\IEEEauthorrefmark{1}}%
\IEEEauthorrefmark{1}Technical University Munich, Germany, 
\IEEEauthorrefmark{2} Siemens AG, Germany \\
\footnotesize{E-Mail:\IEEEauthorrefmark{1}\{ermin.sakic, wolfgang.kellerer\}@tum.de, \IEEEauthorrefmark{2} ermin.sakic@siemens.com}
}

\maketitle

\begin{abstract}
	Distributed Software Defined Networking (SDN) controllers aim to solve the issue of single-point-of-failure and improve the scalability of the control plane. Byzantine and faulty controllers, however, may enforce incorrect configurations and thus endanger the control plane correctness. Multiple Byzantine Fault Tolerance (BFT) approaches relying on Replicated State Machine (RSM) execution have been proposed in the past to cater for this issue. The scalability of such solutions is, however, limited. Additionally, the interplay between progressing the state of the distributed controllers and the consistency of the external reconfigurations of the forwarding devices has not been thoroughly investigated. In this work, we propose an agreement-and-execution group-based approach to increase the overall throughput of a BFT-enabled distributed SDN control plane. We adapt a proven sequencing-based BFT protocol, and introduce two optimized BFT protocols that preserve the uniform agreement, causality and liveness properties. A state-hashing approach which ensures causally ordered switch reconfigurations is proposed, that enables an opportunistic RSM execution without relying on strict sequencing. The proposed designs are implemented and validated for two realistic topologies, a path computation application and a set of KPIs: switch reconfiguration (response) time, signaling overhead, and acceptance rates. We show a clear decrease in the system response time and communication overhead with the proposed models, compared to a state-of-the-art approach.
\end{abstract}


%
\IEEEpeerreviewmaketitle

\section{Introduction and Problem Statement}

Software Defined Networking (SDN) centralizes the decision-making in a dedicated \emph{controller} component. Concepts for achieving crash-fault-tolerance and scalable operation of the controller have been presented in the past \cite{suh2016performance, sakicresponse}. By means of a logical distribution of controller replicas and the state synchronization, the controller instances are able to synchronize the results of their individual computations and come to consistent decisions independent of the instance that handled the client request. However, these approaches are based on weak crash-tolerant algorithms (e.g. RAFT \cite{howard2015raft} and Paxos \cite{lamport2001paxos}) that are unable to cater for malicious and incorrect (e.g., buggy \cite{petramining}) controller decisions that have an individual controller instance fault as a root cause.
Recent works have thus highlighted the importance of deploying Byzantine Fault Tolerance (BFT) protocols for achieving consensus, in scenarios where a subset of controllers is faulty due to a malicious adversary or internal bugs. Realizing a BFT SDN control plane comes with an additional controller deployment overhead, previously shown to range between $2F_M+F_A+1$ \cite{sakicmorph} and $3(F_M+F_A)+1$ \cite{li2014byzantine} controller instances required to tolerate up to $F_M$ strictly Byzantine and $F_A$ fail-crash failures. 

To support stateful controller-based applications (i.e., resource-constrained routing, load-balancing, stateful firewalls), the controllers synchronize their internal state updates. Traditional BFT designs \cite{castro1999practical, li2014byzantine} require active participation of all replicas in the system. Thus, they leverage an RSM approach to handle the client requests, where a majority of controller instances must come to the agreement about the order of the client requests, before subsequently executing them. Finally, the controllers reach consensus on the output of the computation in order to ensure the causality of subsequent decisions. We have identified two issues with this approach.

First, to preserve causality, the non-faulty replicas always participate in all system operations. In the absence of faults, more replicas execute the decision-making requests than required to make progress, thus strongly limiting the execution throughput of the system. Namely, the application execution is handled by each controller instance in the cluster. In heterogeneous environments, where particular controller replicas can be assigned a higher resource set compared to the others, this leads to an under-utilization of fast replicas, as the system progresses at best at the speed of the $\lfloor \frac{|\mathcal{C}|+1}{2} \rfloor + 1$ fastest replica ($|\mathcal{C}|$ is the number of deployed controllers) \cite{sakicresponse}. Second, these BFT implementations rely on reaching a successful agreement about the sequence number mapping for each arriving client request, prior to its actual execution. The agreement phase thus necessarily increases the total processing time of individual requests. We claim that the serialization of requests is a \emph{mean to an end} and that the causality of configurations on individual external devices (i.e., switches) is a sufficient constraint. 

\section{Our contribution}

In this work we make a point that an optimal separation of the controller cluster into \emph{sufficiently-sized} agreement and execution (A\&E) groups leads to an overall higher utilization in request processing. In our approach, faster replicas may be leveraged in the intersection of different A\&E groups, while slower replicas may run at their assigned speed without negatively influencing the faster replicas. To identify the A\&E groups, we extend an existing ILP formulation for controller-switch assignment procedure \cite{sakicmorph}. The solver identifies an A\&E group for each deployed switch element, while maximizing the overlap of the members of different groups. The formulation considers the execution capacity of individual controllers, as well as the switch-controller delays as its constraints. The solver executes during runtime, thus optimizing the assignment upon each discovered Byzantine/fail-crash failure.

To cater for the second issue, we adopt the classical Practical BFT (PBFT) approach \cite{castro1999practical} to realize a distributed sequencer in order to minimize the fail-over time in the case of a leader failure. 
We additionally introduce a group-based variant of this protocol, that leverages the partitioning of the total controller set into multiple A\&E groups.  Finally, in addition to the two \emph{agreement-based} designs above, we present an \emph{opportunistic} protocol design. With the opportunistic approach, successful handling of a client request implies reaching a consensus on a \emph{consistent} device reconfiguration while preserving the \emph{causality} of decisions, subsequent to the actual request handling. We achieve the causality and agreement by reaching consensus: i) on the controller state at the time of application execution; ii) on the actual computed output result (to guarantee the consistency of decisions). 

We have implemented these three BFT protocols and have analyzed the overheads of switch reconfiguration time, the communication overhead and the request acceptances rates. We ran our evaluation for emulated Open vSwitch-based Internet2 and Fat-Tree topologies, comprising up to 34 Open vSwitch instances and up to 13 controllers, while considering a varied number of tolerated Byzantine failures.

\emph{Paper structure}: Sec. \ref{model} introduces the overall system model. Sec. \ref{bftprotocols} details the proposed BFT protocols. Sec. \ref{assignment} discusses the ILP formulation for the optimal controller-switch assignment. Sec. \ref{evaluation} presents the evaluation methodology. Sec. \ref{results} discusses the results. Sec. \ref{relatedwork} summarizes the related work. Sec. \ref{conclusion} concludes this paper.

\newcolumntype{P}[1]{>{\centering\arraybackslash}p{#1}}
\newcolumntype{M}[1]{>{\centering\arraybackslash}m{#1}}

\section{System Model}
\label{model}

In \cite{sakicmorph}, we discussed the often neglected differentiation between state-independent (SIA) and state-dependent (SDA) SDN applications. The SDA require an up-to-date and synchronized application state in order to serve the client requests. In this work, we consider solely the global SDA operations where successfully handled client requests result in stateful write operations to the replicated data-store. The subsequent client request executions that result in new writes to the same state must consider the preceding writes for their correctness. The value of the write operation is determined by an execution of a multi-phase BFT protocol. We distinguish \emph{accepting} and \emph{rejecting} protocol executions. \emph{Rejecting} executions are caused by replicas that interrupt the run because of a missing consensus in one of the protocol \emph{phases} (caused by e.g., conflicting seq. no. proposals, faulty controllers and packet loss). We assume that clients retransmit the requests until a successful execution has been acknowledged by the controllers.

Our SDN architecture is comprised of: i) controllers that individually execute an instance of a BFT process; ii) the switches that implement a comparison mechanism for matching controller configuration messages (as per \cite{sakicmorph}); iii) the clients; iv) a REASSIGNER component that maintains the switch-controller assignments (as per \cite{sakicmorph}). The \emph{request-initiating} clients comprise northbound clients (e.g., applications, administrators) and the switches capable of forwarding the client requests as (OpenFlow) \emph{packet-in} messages to the SDN controllers (e.g., routing, load-balancing requests). The \emph{target clients} represent the configuration targets, e.g., switches that are (re)configured as a result of request handling.

We assume a \emph{fair-loss} link abstraction, where a message (re-)transmitted infinitely often is eventually delivered at the recipient. Packets may be arbitrarily dropped, lost, delayed, duplicated and delivered out of order during any of the BFT protocol phases. The SDN control plane is realized in either in-band or out-of-band manner. Control messages exchanged between the controller, switches and clients are assumed to be signed, thus ensuring: i) the integrity of messages exchanged using the SDN data plane; ii) message forging is impossible. 

State-updates distribution assumes an eventually synchronous model as per \cite{miller2016honey}, where different replicas possess different views of the current configuration state for a limited time duration. Eventually, given an appropriately long quiescent period, all correct replicas converge to the same state. We assume that a bounded number of controllers may exhibit Byzantine behavior and/or fail-crash failures, respectively. 


\section{BFT Consensus Protocols and the Controller-Switch Assignment Methodology}
\label{bftprotocols}
The proposed protocols guarantee the following properties: 
\begin{itemize}
	\item \emph{Uniform Agreement}: When a correct replica commits a particular internal state/switch update (i.e., computes a particular response), all correct replicas eventually commit the same update. 
	\item \emph{Liveness}: All correct replicas eventually finalize the processing of each client request. The resulting run is declared either \emph{accepting} or \emph{rejecting}.
	\item \emph{Causality}: The updates to the controller data-store and the per-switch configuration updates are executed in a causally dependent order. The controller's decision to reconfigure a switch take into account all preceding configurations of that switch. 
\end{itemize}

We assume a deployment of a total of $2F_M+F_A+1$ controllers per agreement and execution (A\&E) group in order to tolerate an upper bound of individual $F_M$ Byzantine and $F_A$ fail-crash controller failures in that particular A\&E group. 

In the remainder of this section, we introduce the three BFT protocols: the agreement-based MPBFT and SBFT protocols, and the opportunistic OBFT (ref. Table \ref{overviewprotocols}).

\definecolor{Gray}{gray}{0.9}
\begin{table}[htb]
	\centering
	\caption{Overview of presented BFT protocols.}
	\begin{tabular}{ M{0.8cm}|M{2.9cm}|M{1.4cm}|M{1.4cm} }
		\hline
		Alg. & Name & Type & No. Rounds \\
		\hline
		\rowcolor{Gray}
		MPBFT & \underline{M}odified \underline{PBFT} & Agreement-based & $2$ \\ 
		SBFT & \underline{S}erialized A Priori \underline{BFT} & Agreement-based & $3$ \\ 
		\rowcolor{Gray}
		OBFT & \underline{O}pportunistic A Posteriori \underline{BFT} & Opportunistic & $2$ \\ 
		\hline
	\end{tabular}
	\label{overviewprotocols}
\end{table}

\subsection{Pre-serialization model MPBFT (agreement-based)}
\label{mpbft}

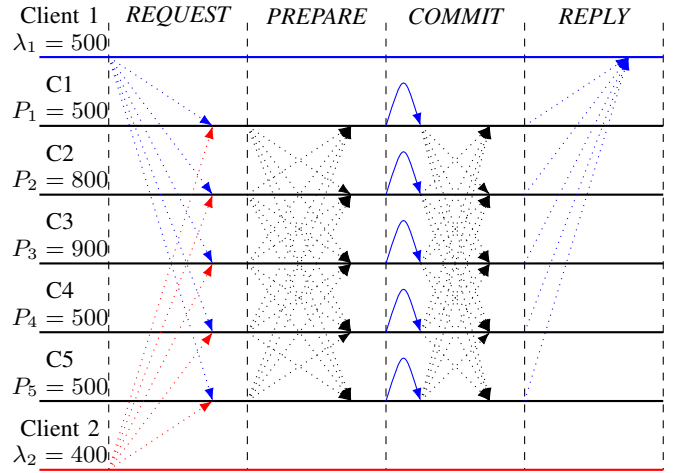
\begin{figure}[htb]
	\begin{center}
		\begin{scaletikzpicturetowidth}{\textwidth/2.1}
			\begin{tikzpicture}[scale=\tikzscale]
				\draw[dashed] (1,5.5) -- (1,-1) ;
				\draw[dashed] (3,5.5) -- (3,-1) ;
				\draw[dashed] (5,5.5) -- (5,-1) ;
				\draw[dashed] (7,5.5) -- (7,-1) ;
				\draw[dashed] (9,5.5) -- (9,-1) ;

				\draw[thick, blue]  (0,5) -- (9,5) ; 
				\draw[thick]  (0,4) -- (9,4) ; 
				\draw[thick]  (0,3) -- (9,3) ; 
				\draw[thick]  (0,2) -- (9,2) ; 
				\draw[thick]  (0,1) -- (9,1) ; 
				\draw[thick]  (0,0) -- (9,0) ; 
				\draw[thick, red]  (0,-1) -- (9,-1) ; 

				\draw[dotted, ->, -Latex, blue] (1,5) --  (2.5,4);
				\draw[dotted, ->, -Latex, blue] (1,5) --  (2.5,3);
				\draw[dotted, ->, -Latex, blue] (1,5) --  (2.5,2);
				\draw[dotted, ->, -Latex, blue] (1,5) --  (2.5,1);
				\draw[dotted, ->, -Latex, blue] (1,5) --  (2.5,0);
				\draw[dotted, ->, -Latex, red] (1,-1) --  (2.5,4);
				\draw[dotted, ->, -Latex, red] (1,-1) --  (2.5,3);
				\draw[dotted, ->, -Latex, red] (1,-1) --  (2.5,2);
				\draw[dotted, ->, -Latex, red] (1,-1) --  (2.5,1);
				\draw[dotted, ->, -Latex, red] (1,-1) --  (2.5,0);

				\draw[dotted, ->, -Latex] (3,4) --  (4.5, 3);
				\draw[dotted, ->, -Latex] (3,4) --  (4.5, 2);
				\draw[dotted, ->, -Latex] (3,4) --  (4.5, 1);
				\draw[dotted, ->, -Latex] (3,4) --  (4.5, 0);

				\draw[dotted, ->, -Latex] (3,2) --  (4.5, 4);
				\draw[dotted, ->, -Latex] (3,2) --  (4.5, 3);
				\draw[dotted, ->, -Latex] (3,2) --  (4.5, 1);
				\draw[dotted, ->, -Latex] (3,2) --  (4.5, 0);

				\draw[dotted, ->, -Latex] (3,3) --  (4.5, 4);
				\draw[dotted, ->, -Latex] (3,3) --  (4.5, 2);
				\draw[dotted, ->, -Latex] (3,3) --  (4.5, 1);
				\draw[dotted, ->, -Latex] (3,3) --  (4.5, 0);

				\draw[dotted, ->, -Latex] (3,1) --  (4.5, 4);
				\draw[dotted, ->, -Latex] (3,1) --  (4.5, 3);
				\draw[dotted, ->, -Latex] (3,1) --  (4.5, 2);
				\draw[dotted, ->, -Latex] (3,1) --  (4.5, 0);

				\draw[dotted, ->, -Latex] (3,0) --  (4.5, 4);
				\draw[dotted, ->, -Latex] (3,0) --  (4.5, 3);
				\draw[dotted, ->, -Latex] (3,0) --  (4.5, 2);
				\draw[dotted, ->, -Latex] (3,0) --  (4.5, 1);

				\draw[blue, ->, -Latex] (5,4) .. controls (5.25, 4.8) .. (5.5,4);
				\draw[dotted, ->, -Latex] (5.5,4) --  (6.5, 3);
				\draw[dotted, ->, -Latex] (5.5,4) --  (6.5, 2);
				\draw[dotted, ->, -Latex] (5.5,4) --  (6.5, 1);
				\draw[dotted, ->, -Latex] (5.5,4) --  (6.5, 0);

				\draw[blue, ->, -Latex] (5,3) .. controls (5.25, 3.8) .. (5.5,3);
				\draw[dotted, ->, -Latex] (5.5,3) --  (6.5, 4);
				\draw[dotted, ->, -Latex] (5.5,3) --  (6.5, 2);
				\draw[dotted, ->, -Latex] (5.5,3) --  (6.5, 1);
				\draw[dotted, ->, -Latex] (5.5,3) --  (6.5, 0);

				\draw[blue, ->, -Latex] (5,2) .. controls (5.25, 2.8) .. (5.5,2);
				\draw[dotted, ->, -Latex] (5.5,2) --  (6.5, 4);
				\draw[dotted, ->, -Latex] (5.5,2) --  (6.5, 3);
				\draw[dotted, ->, -Latex] (5.5,2) --  (6.5, 1);
				\draw[dotted, ->, -Latex] (5.5,2) --  (6.5, 0);

				\draw[blue, ->, -Latex] (5,1) .. controls (5.25, 1.8) .. (5.5,1);
				\draw[dotted, ->, -Latex] (5.5,1) --  (6.5, 4);
				\draw[dotted, ->, -Latex] (5.5,1) --  (6.5, 3);
				\draw[dotted, ->, -Latex] (5.5,1) --  (6.5, 2);
				\draw[dotted, ->, -Latex] (5.5,1) --  (6.5, 0);

				\draw[blue, ->, -Latex] (5,0) .. controls (5.25, 0.8) .. (5.5,0);
				\draw[dotted, ->, -Latex] (5.5,0) --  (6.5, 4);
				\draw[dotted, ->, -Latex] (5.5,0) --  (6.5, 3);
				\draw[dotted, ->, -Latex] (5.5,0) --  (6.5, 2);
				\draw[dotted, ->, -Latex] (5.5,0) --  (6.5, 1);

				\draw[blue, dotted, ->, -Latex] (7,4) --  (8.5, 5);
				\draw[blue, dotted, ->, -Latex] (7,3) --  (8.5, 5);
				\draw[blue, dotted, ->, -Latex] (7,2) --  (8.5, 5);
				\draw[blue, dotted, ->, -Latex] (7,1) --  (8.5, 5);
				\draw[blue, dotted, ->, -Latex] (7,0) --  (8.5, 5);

				\node [ align=center, font=\small] at (0.3, 5.4) {Client 1 \\ $\lambda_1=500$};
				\node [ align=center, font=\small] at (0.3, 4.4) {C1 \\ $P_1 = 500$};
				\node [ align=center, font=\small] at (0.3, 3.4) {C2 \\ $P_2 = 800$};
				\node [ align=center, font=\small] at (0.3, 2.4) {C3 \\ $P_3 = 900$};
				\node [ align=center, font=\small] at (0.3, 1.4) {C4 \\ $P_4 = 500$};
				\node [ align=center, font=\small] at (0.3, 0.4) {C5 \\ $P_5 = 500$};
				\node [ align=center, font=\small] at (0.3, -0.6) {Client 2 \\ $\lambda_2=400$};

				\node [align=center, font=\small] at (2,5.6) {\it{REQUEST}};
				\node [align=center, font=\small] at (4,5.6) {\it{PREPARE}};
				\node [align=center, font=\small] at (6,5.6) {\it{COMMIT}};
				\node [align=center, font=\small] at (8,5.6) {\it{REPLY}};

			\end{tikzpicture}
		\end{scaletikzpicturetowidth}
	\end{center}
	\caption{MPBFT Model: In \texttt{REQUEST} phase, the clients initiate new executions. During \texttt{PREPARE}, the controller replicas agree on the execution order by reaching consensus on the assigned sequence number for the clients' requests. Each controller executes the request in the \texttt{COMMIT} phase. During \texttt{REPLY}, target clients are notified of reconfigurations. Client 2's requests cannot be serviced as a result of a limited processing capacity of the controllers.} 
	\label{fig:mpbft_graphical}
\end{figure}

\emph{Modified PBFT} (MPBFT) imposes a \textbf{single} A\&E group where each active controller replica is tasked with execution of an agreed command. The workflow of MPBFT is visualized in Fig. \ref{fig:mpbft_graphical}. A \emph{request-initiating} client initially invokes its application request to all active controller replicas (\texttt{REQUEST} phase). For each incoming client request, each controller replica assigns a unique sequence number and distributes this sequence number proposal to the other controllers in the cluster (\texttt{PREPARE} phase). The replicas compare the sequence number proposals. If the correct majority of proposals are matching (i.e., the same sequence number is proposed by the majority of correct replicas), successful global agreement has been reached. At the begin of the \texttt{COMMIT} phase, each correct replicas executes the client request. The execution output is subsequently broadcasted by each replica to the remainder of the cluster and the collected output responses are once again compared in all replicas. Each controller deduces the correct majority response and eventually commits the output to its local data-store (i.e., a store of reservations) and finally reports the agreed output to the target clients (\texttt{REPLY} phase). After collecting $F_M+1$ consistent output messages, the \emph{target clients} (e.g., switches) decide to apply the new configuration. 

MPBFT is a variation of PBFT \cite{castro1999practical} that requires no leader and is thus tolerant to individual node failures. Compared to PBFT, we shorten the protocol execution by one round. Whereas PBFT proposes a \texttt{PRE-PREPARE} round, MPBFT skips this round by leveraging a client-initiated atomic multicast execution and a \emph{distributed sequencer}. Namely, each new client request is multicasted to each replica of the system. The replicas propose a new seq. number for the request by incrementing the current counter as per Alg. \ref{distributed-sequencer}. The seq. numbers for new client requests are assigned based on the current state of a local atomic counter. Following an arrival of a new request, the replicas yield the lowest unallocated seq. number value and propose this seq. number to the remaining replicas. After collecting a \emph{sufficient} amount of matching \texttt{PREPARE} messages, all \emph{correct} replicas decide to accept the seq. number contained in the correct majority proposal as the final seq. number for this request. Table \ref{confirmations} summarizes the exact amounts of required matching messages to progress the protocol execution.

If no correct majority vote is achieved during the agreement process on either the sequence number or the computed output, the replicas respond with a rejection status. If sufficient rejection messages are collected, the current execution is cancelled and the run is declared \emph{rejecting}. Concurrent client requests can lead to same sequence numbers being assigned to different requests at different replicas, thus resulting in rejecting runs. 

The execution capacity of MPBFT is limited by the slowest replica in the system. Consider the scenario $F_M =1$, $F_A=0$ depicted in Fig. \ref{fig:mpbft_graphical}. Each controller $C_i$ is able to service request workload up to a capacity of $P_i$ per observation interval. The portrayed system is thus able to service computations up to $max(\sum_{i = 1..N}{\lambda_i}) \leq min(P_i)$, or $500$ requests/interval (imposed by the capacity of $C4$ and $C5$). Thus, Client 1 (with processing requirement of $\lambda_1 = 500$) and Client 2 ($\lambda_2 = 400$) cannot be serviced concurrently. One can alternatively portray the depicted rates as continuous execution workloads. While active participation of $C4$ and $C5$ in the system is unnecessary to tolerate a single Byzantine fault, they are included in execution and signaling and are necessary to progress the system state. 
MPBFT's communication overhead is quadratic (ref. Table \ref{overhead}). With alternative protocol designs SBFT and OBFT, we next leverage the additional execution capacity by partitioning the control plane into multiple A\&E groups. 


\algnewcommand\algorithmicswitch{\textbf{switch}}
\algnewcommand\algorithmiccase{\textbf{case}}
\algnewcommand\algorithmicassert{\texttt{assert}}
\algnewcommand\Assert[1]{\State \algorithmicassert(#1)}%
\algdef{SE}[SWITCH]{Switch}{EndSwitch}[1]{\algorithmicswitch\ #1\ \algorithmicdo}{\algorithmicend\ \algorithmicswitch}%
\algdef{SE}[CASE]{Case}{EndCase}[1]{\algorithmiccase\ #1}{\algorithmicend\ \algorithmiccase}%
\algtext*{EndSwitch}%
\algtext*{EndCase}%

\begin{algorithm}[htb]
	\scriptsize
	\caption{Logical Sequencer: Ordering of client requests}
	\label{distributed-sequencer}
	\hspace*{\algorithmicindent} \textbf{Notation}: \\
	\hspace*{\algorithmicindent} $M_P$ Client request (e.g. flow request)\\
	\hspace*{\algorithmicindent} $M_C$ Replica message (seq. no. proposal) initiated at a remote controller\\
	\hspace*{\algorithmicindent} $\mathcal{C}$ Set of available SDN controllers\\
	\hspace*{\algorithmicindent} $R_{ID}$ Unique client request identifier\\
	\hspace*{\algorithmicindent} $\mathcal{R}_{mappings}$ Mapping of client request ids to unique seq. numbers\\
	\hspace*{\algorithmicindent} $S_{atomic}$ Atomic sequencer that yields the current seq. number\\

	\begin{algorithmic}[1]
		\BState \textbf{upon event} \emph{on-client-request} $<M_P, R_{ID}>$ \textbf{do} 
		\State \emph{...}
		\State proposed\_seq\_no = {\emph{propose\_seq\_no($R_{ID}$)}}
		\State \emph{...}
		\\
		\BState \textbf{upon event} \emph{on-new-replica-sync-update} $<M_C, R_{ID}>$ \textbf{do} 
		\State \emph{...}
		\Switch{PHASE}
		\Case{MPBFT-PREPARE:}
		\State \emph{propose\_seq\_no($R_{ID}$)}
		\EndCase
		\Case{SBFT-PRE-PREPARE:}
		\State \emph{propose\_seq\_no($R_{ID}$)}
		\EndCase
		\EndSwitch
		\State \emph{...}
		\\	
		\Function{propose\_seq\_no($R_{ID}$)}{}
		\If {$R_{ID} \in \mathcal{R}_{mappings}$}
		\State \emph{return} $\mathcal{R}_{mappings}[R_{ID}]$
		\Else
		\While {$S_{atomic} \in \mathcal{R}_{mappings}.values()$}
		\State $S_{atomic} = S_{atomic} + 1$
		\EndWhile
		\State $\mathcal{R}_{mappings}[R_{ID}]$ = $S_{atomic}$ 
		\State \emph{return} $\mathcal{R}_{mappings}[R_{ID}]$
		\EndIf
		\EndFunction
	\end{algorithmic}
\end{algorithm}

\subsection{Pre-serialization model SBFT (agreement-based)}
\label{robft}

\begin{figure} [htb]
	\begin{centering}
		\begin{scaletikzpicturetowidth}{\textwidth/2.1}
			\begin{tikzpicture}[scale=\tikzscale]
				\draw[dashed] (1.2,5.5) -- (1.2,-1) ;
				\draw[dashed] (3,5.5) -- (3,-1) ;
				\draw[dashed] (5,5.5) -- (5,-1) ;
				\draw[dashed] (7,5.5) -- (7,-1) ;
				\draw[dashed] (9,5.5) -- (9,-1) ;
				\draw[dashed] (11,5.5) -- (11,-1) ;

				\draw[thick, blue]  (0,5) -- (11,5) ; 
				\draw[thick]  (0,4) -- (11,4) ; 
				\draw[thick]  (0,3) -- (11,3) ; 
				\draw[thick]  (0,2) -- (11,2) ; 
				\draw[thick]  (0,1) -- (11,1) ; 
				\draw[thick]  (0,0) -- (11,0) ; 
				\draw[thick, red]  (0,-1) -- (11,-1) ; 

				\draw[blue, dotted, ->, -Latex] (1.2,5) --  (2.5,4);
				\draw[blue, dotted, ->, -Latex] (1.2,5) --  (2.5,3);
				\draw[blue, dotted, ->, -Latex] (1.2,5) --  (2.5,2);
				\draw[red, dotted, ->, -Latex] (1.2,-1) --  (2.5,2);
				\draw[red, dotted, ->, -Latex] (1.2,-1) --  (2.5,1);
				\draw[red, dotted, ->, -Latex] (1.2,-1) --  (2.5,0);

				\draw[dotted, ->, -Latex] (3,4) --  (4.8, 3);
				\draw[dotted, ->, -Latex] (3,4) --  (4.8, 2);
				\draw[dotted, ->, -Latex] (3,4) --  (4.8, 1);
				\draw[dotted, ->, -Latex] (3,4) --  (4.8, 0);

				\draw[dotted, ->, -Latex] (3,3) --  (4.8, 4);
				\draw[dotted, ->, -Latex] (3,3) --  (4.8, 2);
				\draw[dotted, ->, -Latex] (3,3) --  (4.8, 1);
				\draw[dotted, ->, -Latex] (3,3) --  (4.8, 0);

				\draw[dotted, ->, -Latex] (3,2) --  (4.8, 4);
				\draw[dotted, ->, -Latex] (3,2) --  (4.8, 3);
				\draw[dotted, ->, -Latex] (3,2) --  (4.8, 1);
				\draw[dotted, ->, -Latex] (3,2) --  (4.8, 0);

				\draw[dotted, ->, -Latex] (3,1) --  (4.8, 4);
				\draw[dotted, ->, -Latex] (3,1) --  (4.8, 3);
				\draw[dotted, ->, -Latex] (3,1) --  (4.8, 2);
				\draw[dotted, ->, -Latex] (3,1) --  (4.8, 0);

				\draw[dotted, ->, -Latex] (3,0) --  (4.8, 4);
				\draw[dotted, ->, -Latex] (3,0) --  (4.8, 3);
				\draw[dotted, ->, -Latex] (3,0) --  (4.8, 2);
				\draw[dotted, ->, -Latex] (3,0) --  (4.8, 1);

				\draw[dotted, ->, -Latex] (5,4) --  (6.8, 3);
				\draw[dotted, ->, -Latex] (5,4) --  (6.8, 2);
				\draw[dotted, ->, -Latex] (5,4) --  (6.8, 1);
				\draw[dotted, ->, -Latex] (5,4) --  (6.8, 0);

				\draw[dotted, ->, -Latex] (5,3) --  (6.8, 4);
				\draw[dotted, ->, -Latex] (5,3) --  (6.8, 2);
				\draw[dotted, ->, -Latex] (5,3) --  (6.8, 1);
				\draw[dotted, ->, -Latex] (5,3) --  (6.8, 0);

				\draw[dotted, ->, -Latex] (5,2) --  (6.8, 4);
				\draw[dotted, ->, -Latex] (5,2) --  (6.8, 3);
				\draw[dotted, ->, -Latex] (5,2) --  (6.8, 1);
				\draw[dotted, ->, -Latex] (5,2) --  (6.8, 0);

				\draw[dotted, ->, -Latex] (5,1) --  (6.8, 4);
				\draw[dotted, ->, -Latex] (5,1) --  (6.8, 3);
				\draw[dotted, ->, -Latex] (5,1) --  (6.8, 2);
				\draw[dotted, ->, -Latex] (5,1) --  (6.8, 0);

				\draw[dotted, ->, -Latex] (5,0) --  (6.8, 4);
				\draw[dotted, ->, -Latex] (5,0) --  (6.8, 3);
				\draw[dotted, ->, -Latex] (5,0) --  (6.8, 2);
				\draw[dotted, ->, -Latex] (5,0) --  (6.8, 1);

				\draw[blue, ->, -Latex] (7,4) .. controls (7.25, 4.8) .. (7.5,4);
				\draw[dotted, ->, -Latex] (7.5,4) --  (8.8, 3);
				\draw[dotted, ->, -Latex] (7.5,4) --  (8.8, 2);
				\draw[dotted, ->, -Latex] (7.5,4) --  (8.8, 1);
				\draw[dotted, ->, -Latex] (7.5,4) --  (8.8, 0);

				\draw[blue, ->, -Latex] (7,3) .. controls (7.25, 3.8) .. (7.5,3);
				\draw[dotted, ->, -Latex] (7.5,3) --  (8.8, 4);
				\draw[dotted, ->, -Latex] (7.5,3) --  (8.8, 2);
				\draw[dotted, ->, -Latex] (7.5,3) --  (8.8, 1);
				\draw[dotted, ->, -Latex] (7.5,3) --  (8.8, 0);

				\draw[blue, ->, -Latex] (7,2) .. controls (7.1, 2.9) and (7.4, 2.9) .. (7.5,2);
				\draw[red, ->, -Latex] (7,2) .. controls (7.25, 2.6) .. (7.48,2);
				\draw[dotted, ->, -Latex] (7.5,2) --  (8.8, 4);
				\draw[dotted, ->, -Latex] (7.5,2) --  (8.8, 3);
				\draw[dotted, ->, -Latex] (7.5,2) --  (8.8, 1);
				\draw[dotted, ->, -Latex] (7.5,2) --  (8.8, 0);

				\draw[red, ->, -Latex] (7,1) .. controls (7.25, 1.8) .. (7.5,1);
				\draw[dotted, ->, -Latex] (7.5,1) --  (8.8, 4);
				\draw[dotted, ->, -Latex] (7.5,1) --  (8.8, 3);
				\draw[dotted, ->, -Latex] (7.5,1) --  (8.8, 2);
				\draw[dotted, ->, -Latex] (7.5,1) --  (8.8, 0);

				\draw[red, ->, -Latex] (7,0) .. controls (7.25, 0.8) .. (7.5,0);
				\draw[dotted, ->, -Latex] (7.5,0) --  (8.8, 4);
				\draw[dotted, ->, -Latex] (7.5,0) --  (8.8, 3);
				\draw[dotted, ->, -Latex] (7.5,0) --  (8.8, 2);
				\draw[dotted, ->, -Latex] (7.5,0) --  (8.8, 1);

				\draw[blue, dotted, ->, -Latex] (9,4) --  (10.5, 5);
				\draw[blue, dotted, ->, -Latex] (9,3) --  (10.5, 5);
				\draw[blue, dotted, ->, -Latex] (9,2) --  (10.5, 5);

				\draw[red, dotted, ->, -Latex] (9,2) --  (10.5, -1);
				\draw[red, dotted, ->, -Latex] (9,1) --  (10.5, -1);
				\draw[red, dotted, ->, -Latex] (9,0) --  (10.5, -1);

				\draw[blue, rounded corners, dashed, thick] (-0.7,1.9) rectangle (11.2,4.9);
				\draw[red, rounded corners, dashed, thick] (-0.7,-0.15) rectangle (11.2,2.9);

				\node [align=center, font=\small] at (0.3, 5.5) {Client 1\\$\lambda_1=500$};
				\node [align=center, font=\small] at (0.3, 4.5) {C1\\$P_1 = 500$};
				\node [align=center, font=\small] at (0.3, 3.5) {C2\\$P_2 = 800$};
				\node [align=center, font=\small] at (0.3, 2.5) {C3\\$P_3 = 900$};
				\node [align=center, font=\small] at (0.3, 1.5) {C4\\$P_4 = 500$};
				\node [align=center, font=\small] at (0.3, 0.5) {C5\\$P_5 = 500$};
				\node [align=center, font=\small] at (0.3, -0.5) {Client 2\\$\lambda_2=400$};

				\node [ align=center, font=\small] at (2,6) {\it{REQUEST}};
				\node [ align=center, font=\small] at (4,6) {\it{PRE-}\\\it{PREPARE}};
				\node [ align=center, font=\small] at (6,6) {\it{PREPARE}};
				\node [ align=center, font=\small] at (8,6) {\it{COMMIT}};
				\node [ align=center, font=\small] at (10,6) {\it{REPLY}};

			\end{tikzpicture}
		\end{scaletikzpicturetowidth}
	\end{centering}
	\caption{SBFT Model: Compared to MPBFT, SBFT allows for more efficient allocation of execution resources, since execution is separated into multiple A\&E groups. This comes with an overhead of a \texttt{PRE-PREPARE} step, required to reach consensus on the sequence number allocated to the request.} 
	\label{fig:robft_graphical}
\end{figure}
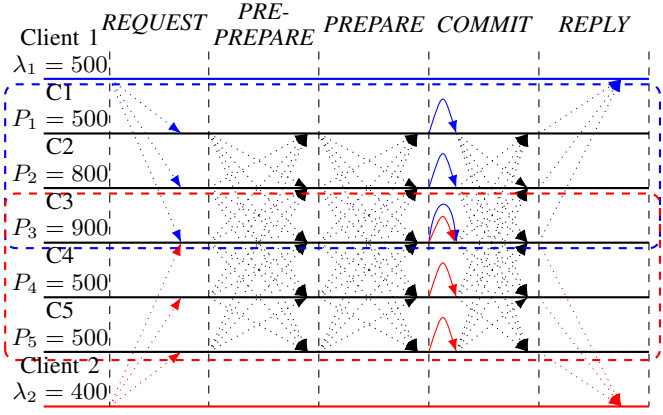

With \emph{Serialized A Priori BFT} (SBFT), agreement and execution processes are administered by multiple A\&E groups. We assign for each request-initiating client (i.e., a northbound application, an edge switch) an A\&E group according to the algorithm presented in Sec. \ref{assignment}. To tolerate $F_M$ Byzantine and $F_A$ fail-crash failures in the scope of a single A\&E group, each group must comprise $2F_M+F_A+1$ controllers. Multiple execution groups can process the client requests concurrently. SBFT design is depicted in Fig. \ref{fig:robft_graphical}. Compared to MPBFT, SBFT introduces the \texttt{PRE-PREPARE} step, where the replicas belonging to the A\&E group propose and subsequently notify the remainder of the replicas of an assigned sequence number. In an \emph{accepting} run, the group replicas collect the responses in the \texttt{PREPARE} phase and reach consensus by collecting $\lceil \frac{|\mathcal{C}|+F_M+1}{2} \rceil$ matching sequence number proposals. Finally, the replicas of the A\&E group execute the request in the \texttt{COMMIT} phase and broadcast the response to all remaining replicas. If $F_M+1$ matching outputs are received, the replicas apply the internal state reconfiguration and notify the target clients of the final result during \texttt{REPLY}. The communication overhead of SBFT is bounded $\mathcal{O}(3|\mathcal{A}||\mathcal{C}|)$, and grows linearly for a fixed A\&E group size. 

\emph{Causality}: To ensure that the causality property holds in MPBFT and SBFT, the controllers execute the sequenced request in order agreed during \texttt{PREPARE}. The replicas execute the \texttt{COMMIT} phase only if the outputs (i.e., the added reservations) for the preceding requests were seen by the executing replica. Thus, before handling subsequent requests, the status of preceding runs (\emph{accepting}/\emph{rejecting}) must be determined. 

\subsection{Post-negotiation model OBFT (opportunistic)}
\label{oobft}

\begin{figure} [htb]
	\begin{center}
		\begin{scaletikzpicturetowidth}{\textwidth/2.1}
			\begin{tikzpicture}[scale=\tikzscale]
				\draw[dashed] (1.1,5.5) -- (1.1,-1) ;
				\draw[dashed] (3,5.5) -- (3,-1) ;
				\draw[dashed] (5,5.5) -- (5,-1) ;
				\draw[dashed] (7,5.5) -- (7,-1) ;
				\draw[dashed] (9,5.5) -- (9,-1) ;

				\draw[thick, blue]  (0,5) -- (9,5) ; 
				\draw[thick]  (0,4) -- (9,4) ; 
				\draw[thick]  (0,3) -- (9,3) ; 
				\draw[thick]  (0,2) -- (9,2) ; 
				\draw[thick]  (0,1) -- (9,1) ; 
				\draw[thick]  (0,0) -- (9,0) ; 
				\draw[thick, red]  (0,-1) -- (9,-1) ; 

				\draw[blue, dotted, ->, -Latex] (1,5) --  (2.5,4);
				\draw[blue, dotted, ->, -Latex] (1,5) --  (2.5,3);
				\draw[blue, dotted, ->, -Latex] (1,5) --  (2.5,2);
				\draw[red, dotted, ->, -Latex] (1,-1) --  (2.5,2);
				\draw[red, dotted, ->, -Latex] (1,-1) --  (2.5,1);
				\draw[red, dotted, ->, -Latex] (1,-1) --  (2.5,0);

				\draw[blue, ->, -Latex] (3,4) .. controls (3.25, 4.8) .. (3.5,4);
				\draw[dotted, ->, -Latex] (3.5,4) --  (4.8, 3);
				\draw[dotted, ->, -Latex] (3.5,4) --  (4.8, 2);
				\draw[dotted, ->, -Latex] (3.5,4) --  (4.8, 1);
				\draw[dotted, ->, -Latex] (3.5,4) --  (4.8, 0);

				\draw[blue, ->, -Latex] (3,3) .. controls (3.25, 3.8) .. (3.5,3);
				\draw[dotted, ->, -Latex] (3.5,3) --  (4.8, 4);
				\draw[dotted, ->, -Latex] (3.5,3) --  (4.8, 2);
				\draw[dotted, ->, -Latex] (3.5,3) --  (4.8, 1);
				\draw[dotted, ->, -Latex] (3.5,3) --  (4.8, 0);

				\draw[blue, ->, -Latex] (3,2) .. controls (3.1, 2.9) and (3.4, 2.9) .. (3.5,2);
				\draw[red, ->, -Latex]  (3,2) .. controls (3.25, 2.6) .. (3.48,2);
				\draw[dotted, ->, -Latex] (3.5,2) --  (4.8, 4);
				\draw[dotted, ->, -Latex] (3.5,2) --  (4.8, 3);
				\draw[dotted, ->, -Latex] (3.5,2) --  (4.8, 1);
				\draw[dotted, ->, -Latex] (3.5,2) --  (4.8, 0);

				\draw[red, ->, -Latex] (3,1) .. controls (3.25, 1.8) .. (3.5,1);
				\draw[dotted, ->, -Latex] (3.5,1) --  (4.8, 4);
				\draw[dotted, ->, -Latex] (3.5,1) --  (4.8, 3);
				\draw[dotted, ->, -Latex] (3.5,1) --  (4.8, 2);
				\draw[dotted, ->, -Latex] (3.5,1) --  (4.8, 0);

				\draw[red, ->, -Latex] (3,0) .. controls (3.25, 0.8) .. (3.5,0);
				\draw[dotted, ->, -Latex] (3.5,0) --  (4.8, 4);
				\draw[dotted, ->, -Latex] (3.5,0) --  (4.8, 3);
				\draw[dotted, ->, -Latex] (3.5,0) --  (4.8, 2);
				\draw[dotted, ->, -Latex] (3.5,0) --  (4.8, 1);

				\draw[dotted, ->, -Latex] (5,4) --  (6.5, 3);
				\draw[dotted, ->, -Latex] (5,4) --  (6.5, 2);
				\draw[dotted, ->, -Latex] (5,4) --  (6.5, 1);
				\draw[dotted, ->, -Latex] (5,4) --  (6.5, 0);

				\draw[dotted, ->, -Latex] (5,3) --  (6.5, 4);
				\draw[dotted, ->, -Latex] (5,3) --  (6.5, 2);
				\draw[dotted, ->, -Latex] (5,3) --  (6.5, 1);
				\draw[dotted, ->, -Latex] (5,3) --  (6.5, 0);

				\draw[dotted, ->, -Latex] (5,2) --  (6.5, 4);
				\draw[dotted, ->, -Latex] (5,2) --  (6.5, 3);
				\draw[dotted, ->, -Latex] (5,2) --  (6.5, 1);
				\draw[dotted, ->, -Latex] (5,2) --  (6.5, 0);

				\draw[dotted, ->, -Latex] (5,1) --  (6.5, 4);
				\draw[dotted, ->, -Latex] (5,1) --  (6.5, 3);
				\draw[dotted, ->, -Latex] (5,1) --  (6.5, 2);
				\draw[dotted, ->, -Latex] (5,1) --  (6.5, 0);

				\draw[dotted, ->, -Latex] (5,0) --  (6.5, 4);
				\draw[dotted, ->, -Latex] (5,0) --  (6.5, 3);
				\draw[dotted, ->, -Latex] (5,0) --  (6.5, 2);
				\draw[dotted, ->, -Latex] (5,0) --  (6.5, 1);

				\draw[blue, dotted, ->, -Latex] (7,4) --  (8.5, 5);
				\draw[blue, dotted, ->, -Latex] (7,3) --  (8.5, 5);
				\draw[blue, dotted, ->, -Latex] (7,2) --  (8.5, 5);

				\draw[red, dotted, ->, -Latex] (7,2) --  (8.5, -1);
				\draw[red, dotted, ->, -Latex] (7,1) --  (8.5, -1);
				\draw[red, dotted, ->, -Latex] (7,0) --  (8.5, -1);

				\draw[blue, rounded corners, dashed, thick] (-0.6,1.8) rectangle (9.3,4.8);
				\draw[red, rounded corners, dashed, thick] (-0.6,-0.16) rectangle (9.3,2.8);

				\node [ align=center, font=\small] at (0.3, 5.4) {Client 1 \\ $\lambda_1=500$};
				\node [ align=center, font=\small] at (0.3, 4.4) {C1 \\ $P_1 = 500$};
				\node [ align=center, font=\small] at (0.3, 3.4) {C2 \\ $P_2 = 800$};
				\node [ align=center, font=\small] at (0.3, 2.4) {C3 \\ $P_3 = 900$};
				\node [ align=center, font=\small] at (0.3, 1.4) {C4 \\ $P_4 = 500$};
				\node [ align=center, font=\small] at (0.3, 0.4) {C5 \\ $P_5 = 500$};
				\node [ align=center, font=\small] at (0.3, -0.6) {Client 2 \\ $\lambda_2=400$};

				\node [ align=center, font=\small] at (2,5.6) {\it{REQUEST}};
				\node [ align=center, font=\small] at (4,5.6) {\it{COMMIT}};
				\node [ align=center, font=\small] at (6,5.6) {\it{PRE-REPLY}};
				\node [ align=center, font=\small] at (8,5.6) {\it{REPLY}};
			\end{tikzpicture}
		\end{scaletikzpicturetowidth}
	\end{center}
	\caption{OBFT model: An opportunistic protocol variation, where A\&E group members execute their clients' requests prior to the distribution of reference state configurations based on which the computations were executed. The internal controller state and the target clients are updated only if the consensus on the reference state configurations could be reached for the \emph{correct} majority of global controller instances (ref. Table \ref{confirmations}).} 
	\label{fig:oobft_graphical}
\end{figure}
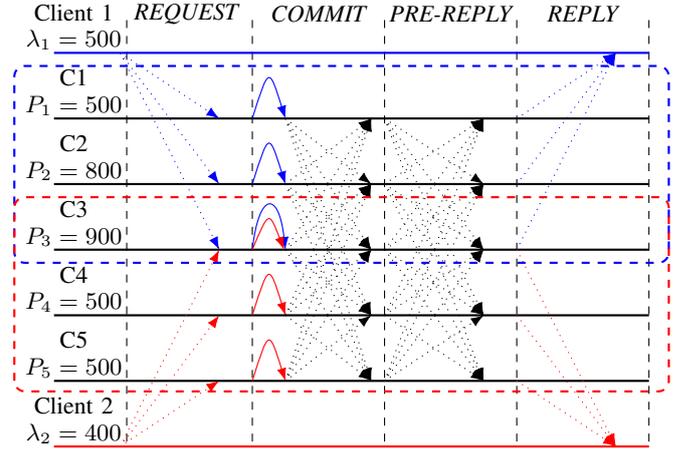

\emph{Opportunistic A Posteriori BFT} (OBFT) is a speculative take on SBFT, where computations of the client requests execute prior to reaching consensus about the computed output values. A global sequencer is not used in OBFT and thus \texttt{PRE-PREPARE} and \texttt{PREPARE} phases are omitted. Instead, each replica maintains the hashes of current switch configurations, as well as a state array containing the hashes of the configurations of the switches at the \underline{t}ime \underline{o}f \underline{r}equest exe\underline{c}utions (\emph{TORC} hashes). Following the output computation in the \texttt{COMMIT} phase, the replicas come to consensus about the updated switch state in the \texttt{PRE-REPLY} phase. This workflow is depicted in Fig. \ref{fig:oobft_graphical}.

In contrast to MPBFT and SBFT, in their \texttt{COMMIT} phase, the replicas belonging to the same A\&E group compute the outputs, and in addition to the computed response outputs, they broadcast the hash arrays denoting their view of the target clients' configurations. Each accepting replica \emph{that is not part of the serving A\&E group} evaluates its actual current local view of the switch states, and \emph{iff}: i) $F_M+1$ matching output values have been computed by the A\&E replicas; and b) their current view of switch configuration hashes is matching with those of the A\&E replicas; they answer with an \emph{accepting} status. The execution replicas (belonging to the A\&E group), instead compare the proposed hash array with their local TORC hashes for the target client (i.e., target switches) and notify other A\&E replicas of their status. If \emph{sufficient} (ref. Table \ref{confirmations}) positive confirmations have been collected at the end of \texttt{PRE-REPLY} phase, each active controller internally commits the output proposed by the correct majority of the A\&E group. The A\&E group members then notify the configuration targets of the agreed output in \texttt{REPLY} phase. OBFT's comm. overhead is quadratic and grows with $|\mathcal{C}|$.

\begin{algorithm}[htb]
	\caption{Hash comparison in the OBFT-COMMIT phase}
	\label{oobft-hash-comparator}
	\footnotesize
	\hspace*{\algorithmicindent} \textbf{Notation}: \\
	\hspace*{\algorithmicindent} $R_{ID}$ Unique client request identifier\\
	\hspace*{\algorithmicindent} $\mathcal{HV}$ Current configuration hashes for the switches\\
	\hspace*{\algorithmicindent} $\mathcal{HVC[R_{ID}]}$ Switches' config. hashes prior request computation (TORC)\\
	\hspace*{\algorithmicindent} \emph{find-path()} An exemplary SDN application logic operation\\ 
	\hspace*{\algorithmicindent} \emph{consensus()} Returns consensus message according to the number of minimum required confirmations (ref. Table \ref{confirmations}) \\ 
	\hspace*{\algorithmicindent} $m_{R_{ID}}^{C}$ A COMMIT message for round $R_{ID}$\\
	\hspace*{\algorithmicindent} $M_{R_{ID}}^{C}$ Set of buffered COMMIT messages for $R_{ID}$ \\

	\begin{algorithmic}[1]
		\Procedure{Handle new client request}{}
		\BState \textbf{upon event} \emph{on-received-client-request} ($CL_{R_{ID}}$) \textbf{do} 
		\State $\mathcal{R} =$ \emph{find-path}($CL_{R_{ID}}$.routing\_request)
		\ForAll {SW $\in \mathcal{R}$}
		\State current-hash[SW]  = \emph{hash}(SW.state)
		\EndFor		
		\State $m_{R_{ID}}^{C}$.hash, $m_{R_{ID}}^{C}$.path  = current-hash, $\mathcal{R}$
		\State \emph{broadcast-to-cluster-members($m_{R_{ID}}^{C}$)}
		\EndProcedure
		\\
		\Procedure{Handle incoming COMMIT message}{}
		\BState \textbf{upon event} \emph{new-replica-sync-message}($m_{R_{ID}}^{C}$) \textbf{do} 
		\State $P_{C}^{R_{ID}}$ = \emph{consensus}($M_{R_{ID}}^{C}$, $<$val, state-hash-array$>$)

		\State	\emph{on-init-obft-pre-reply($P_{C}^{R_{ID}}$,\emph{inline-with-replica-view($P_{C}^{R_{ID}}$)})}
		\EndProcedure
		\\
		\Function{inline-with-replica-view($P_{C}^{R_{ID}}$)}{}
		\ForAll{SW $\in  P_{C}^{R_{ID}}$.path}
		\If {$\mathcal{HV}[$SW$] == P_{C}^{R_{ID}}.hash[$SW$]$}
		\State \emph{pass ()}
		\ElsIf {$\mathcal{HVC}[R_{ID}][$SW$]==P_{C}^{R_{ID}}.hash[$SW$]$}
		\State \emph{pass ()}
		\EndIf
		\State \textbf{else} \emph{return} (REJECT)
		\EndFor
		\State \emph{return} (ACCEPT)
		\EndFunction
	\end{algorithmic}
\end{algorithm}

\begin{table}[htb]
	\centering
	\caption{Notation used in Tables \ref{confirmations}, \ref{overhead} and \ref{ilpconstraints}.}
	\begin{tabular}{ M{1.2cm}|M{6.7cm} }
		\hline
		Symbol & Meaning \\
		\hline
		\rowcolor{Gray}
		$ \mathcal{C} $ & Set of active controllers in the system \\ 
		$F_M$ & No. of tolerated Byzantine faults in a single A\&E group\\ 
		\rowcolor{Gray}
		$ Req(t) $ & Time-variant no. of controllers \cite{sakicmorph} that must be assigned to each switch, to tolerate the Byzantine failures \\
		$ \mathcal{S} $ & Set of switches in the system \\ 
		\rowcolor{Gray}
		$ P_{C_i} $ & Total available controller $C_i$'s capacity. \\ 
		$ L_{CL_k}, L_{S_j} $ & Request processing load stemming from the northbound client $CL_k$ and edge switch $S_j$, respectively. \\ 
		\rowcolor{Gray}
		$D_{C,S} $ & Max. tolerable delay for controller-switch communication. \\ 
		$\mathcal{A}$ & Controller replicas belonging to a single A\&E group\\ 
		\rowcolor{Gray}
		$|\mathcal{M}_{agr}|$ & Sum of the tolerated Byzantine failures and the majority of correct replicas per A\&E group: $\lceil \frac{|\mathcal{A}|+F_M+1}{2} \rceil$ \\ 
		$|\mathcal{M}_{glob}|$ & Sum of the tolerated Byzantine replicas and the majority of all correct active replicas: $\lceil \frac{|\mathcal{C}|+F_M+1}{2} \rceil$ \\ 
		\rowcolor{Gray}
		$\texttt{CMP}$ & Comp. overhead of executing the packet comparison \\ 
		$\texttt{E}$ & Comp. overhead of executing SDN application operation\\ 
		\hline
	\end{tabular}
	\label{notation}
\end{table}

\begin{table}[htb]
	\centering
	\caption{The amount of matching messages required to reach consensus in the respective protocol phase (worst-case).}
	\begin{tabular}{ M{1cm}|M{1cm}|M{1cm}|M{1cm}|M{1cm}|M{1cm} }
		\hline
		Algorithm & PRE-PREPARE & PREPARE & COMMIT & PRE-REPLY & REPLY \\
		\rowcolor{Gray}
		MPBFT & N/A & $|M_{glob}|$ & $F_M+1$ & N/A & $F_M+1$ \\ 
		SBFT & $|M_{agr}|$ & $|M_{glob}|$ & $F_M+1$ & N/A & $F_M+1$ \\ 
		\rowcolor{Gray}
		OBFT & N/A & N/A & $|M_{agr}|$ & $|M_{glob}|$ & $F_M+1$ \\
		\hline
	\end{tabular}
	\label{confirmations}
\end{table}

\begin{table}[htb]
	\centering
	\caption{Computational and communication overhead of the introduced BFT protocols.}
	\begin{tabular}{ P{0.8cm}|P{3.3cm}|P{3.4cm}  }
		\hline
		Alg. & Computational Overhead & Communication Overhead \\
		\hline
		\rowcolor{Gray}
		MPBFT & $\mathcal{O}(2|\mathcal{C}| \texttt{CMP} + |\mathcal{C}| \texttt{E}$) & $\mathcal{O}(2|\mathcal{C}||\mathcal{C}|)$  \\ 
		SBFT & $\mathcal{O}(\texttt{CMP} (2|\mathcal{C}| + |\mathcal{A}|) + |\mathcal{A}| E$) & $\mathcal{O}(3|\mathcal{A}||\mathcal{C}|)$ \\ 
		\rowcolor{Gray}
		OBFT & $\mathcal{O}(2|\mathcal{C}| \texttt{CMP} + |\mathcal{A}| E$) & $\mathcal{O}(|\mathcal{A}|(|\mathcal{C}|+1) + |\mathcal{C}|(|\mathcal{C}|-1))$ \\ 
		\hline
	\end{tabular}
	\label{overhead}
\end{table}

\subsection{Dynamic Controller-Switch (Re)Assignment Procedure}
\label{assignment}

In our design, each request-initiating client (i.e., a northbound client or a switch) is assigned a unique controller agreement and execution (A\&E) group. Groups assigned to different switches are allowed to partially or fully overlap. Only the assigned controllers are required to contact the target clients and apply reconfigurations. Similarly, only these controllers are contacted by the request-initiating clients with new application requests. Our ILP formulation of the assignment problem aims to minimize the total overlap between the members of the active A\&E groups, so to minimize the synchronization delay during the consensus executions. The proposed reassignment mechanism, the objective function and the constraints extend the formulation presented in \cite{sakicmorph}. For brevity, we do not discuss each constraint in detail here, but refer the reader to the summary in Table \ref{ilpconstraints} and \cite{sakicmorph}. The procedure is executed once at the system startup and dynamically during runtime, on each detected controller failure. 

For each switch $S_i$ we can derive a bitstring $R_{S_i}$ comprised of \emph{ones} for replicas actively assigned to $S_i$ and \emph{zeros} for the unassigned replicas. We then formalize the objective function:

\begin{equation}
	min \sum_{S_j\in\mathcal{S}}{\sum_{S_i\in\mathcal{S},S_i \ne S_j}{HD(R_{S_j}, R_{S_i})}}
\end{equation}

where $HD(R_{S_j}, R_{S_i})$ denotes the Hamming distance between the assignment bitstrings for $S_j$ and $S_i$. Combined with the adapted \emph{minimum assignment} constraint depicted in Table \ref{ilpconstraints}, we ensure the building of minimum-sized A\&E groups that fulfill the capacity and delay constraints of the clients.


\begin{table}[htb]
	\centering
	\caption{Constraints used in building the A\&E groups.}
	\begin{tabular}{ M{2.3cm}|M{6cm} }
		\hline
		Constraint & Formulation \\
		\hline
		\rowcolor{Gray}
		Min. Assignment & \pbox{7cm}{$\sum\limits_{C_i\in\mathcal{C}}{A_{C_i,S_j}} == Req(t), \forall S_j \in \mathcal{S}$}  \\ 
		Unique Assignment & \pbox{7cm}{$A_{C_i,S_j} \leq 1, \forall C_i \in \mathcal{C}, S_j \in \mathcal{S}$}\\ 
		\rowcolor{Gray}
		Bounded Capacity & \pbox{7cm}{$\sum\limits_{S_j\in\mathcal{S}}{A_{C_i,S_j}*L_{S_j}} \leq$ \\ $P_{C_i} - \sum\limits_{CL_k\in\mathcal{CL}}{L_{CL_k}}, \forall C_i \in \mathcal{C}$} \\ 
		Delay Bounds & \pbox{7cm}{$A_{C_i,S_j}*d_{C_i, S_j} \leq D_{C,S}, \forall C_i \in \mathcal{C}, S_j\in\mathcal{S}$}\\
		\hline
	\end{tabular}
	\label{ilpconstraints}
\end{table}


\section{Evaluation}
\label{evaluation}

To evaluate the different BFT protocols, we realized a centralized path computation application that executes in each of the deployed controller replicas. Based on the sequence and current state of link reservations, the routing algorithm leverages Dijkstra algorithm to choose the optimal (cheapest) path w.r.t. bandwidth resource consumption, and thus implicitly load-balances the embedded flows in the given topology. The BFT protocol executions take the source-destination pair and the required bandwidth as an input for the service request. Subsequently, the protocol computes the optimal path in the \texttt{COMMIT} phase and notifies the switches on the path of new reservation in the \texttt{REPLY} phase. To evaluate the designs of all three protocols, we consider the following performance metrics: i) time required to apply a new switch reconfiguration, measured from the time of a client request arrival until the confirmation of the last switch reconfiguration; ii) the acceptance rate for the new arrivals; ii) the total communication overhead. 

To validate our claims in a realistic environment, we have emulated the Internet2 topology, as well as a fat-tree data-center topology, encompassing 34 and 20 switches, respectively. The controllers in the Internet2 scenario were placed so to maximize the system coverage against failures as per \cite{sakicmorph, hock2014poco}. The controllers of the fat-tree topology were placed on the leaf-nodes as per \cite{sakicmorph, huang2017dynamic}. The state synchronization between the controllers and the resulting switch reconfigurations occur in in-band control mode. To provide for realistic delay emulation, we derive the link distances from the publicly available geographical Internet2 data\footnote{Internet2 topological data (provided by POCO project) - \url{https://github.com/lsinfo3/poco/tree/master/topologies}} and inject the propagation delays using Linux's \emph{tc} tool. A single client was placed at each switch of the Internet2 topology, while two clients were placed at each leaf-switch of the fat-tree topology. The arrivals for incoming service requests are modeled using n.e.d. \cite{huang2017dynamic}. 

To generate the hashes for per-switch configuration state (ref. Sec. \ref{oobft}), we used Python's \emph{hashlib} implementation and the SHA256 secure hash algorithm, defined in FIPS 180-2 \cite{sha}. We used Gurobi to solve the ILP formulated in Sec. \ref{assignment}. The measurements were executed on a commodity PC equipped with AMD Ryzen 1600 CPU and 32 GB RAM. 


\section{Discussion}
\label{results}

\begin{figure}[htb]
	\centering
	\subfloat[Fat-Tree topology]{
		\centering
		\includegraphics[width=0.49\textwidth]{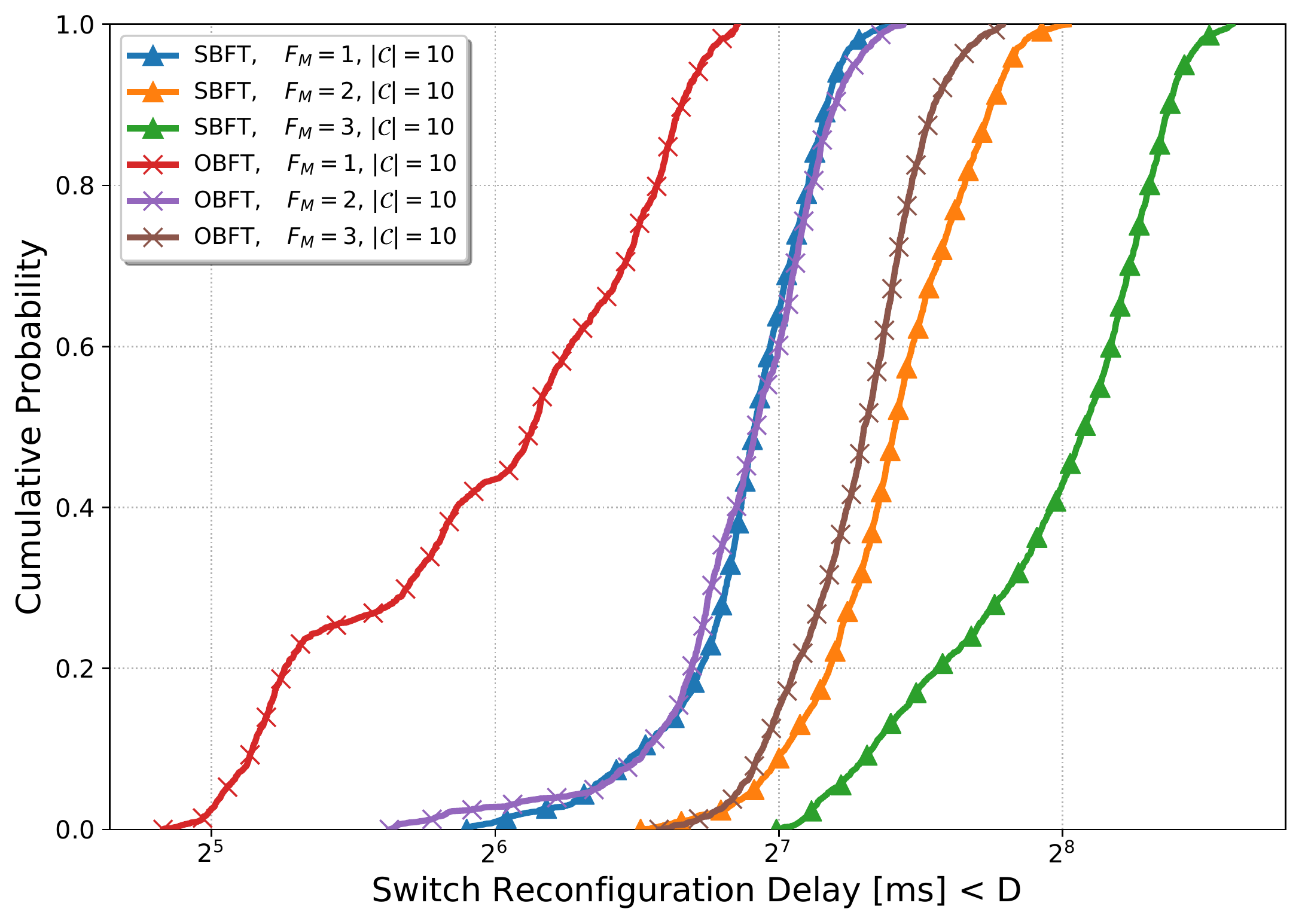}
		\label{fig1:fattree}
	}
	\newline
	\subfloat[Internet2 topology]{
		\centering
		\includegraphics[width=0.49\textwidth]{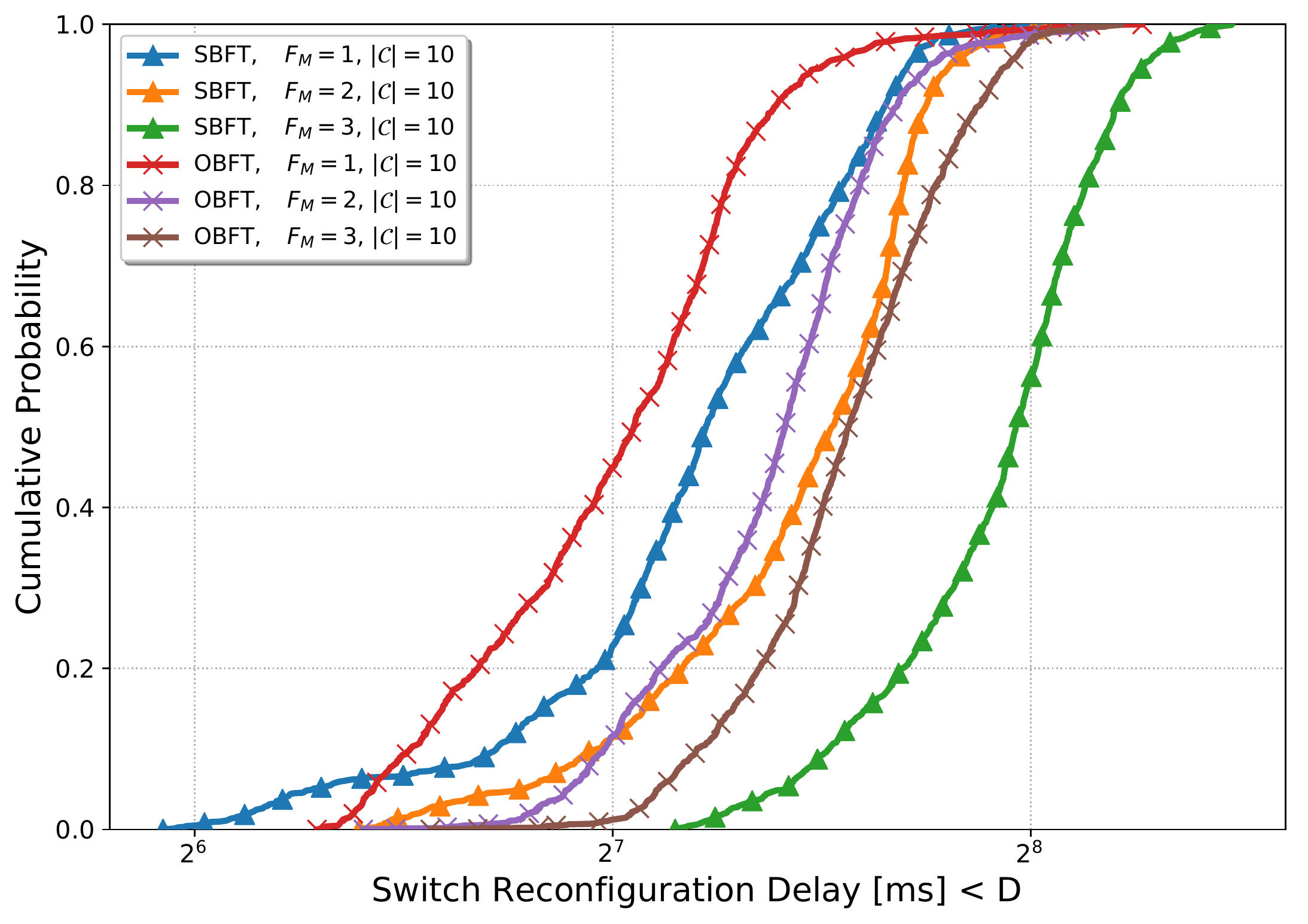}
		\label{fig1:internet2}
	}
	\caption{Total accumulated switch reconfiguration (system response) time for varied sizes of A\&E groups for max. tolerated Byzantine failures $F_M = [1..3]$,  $F_A=0$ and a fixed total number of active controllers $|\mathcal{C}|=[10]$. OBFT shows dominantly lower commit delays in all depicted scenarios.}
	\label{fig:c2sdelaymetrics1}
\end{figure}

\subsubsection{Total reconfiguration time for the internal controller and the switch state} Fig. \ref{fig1:fattree} and Fig. \ref{fig1:internet2} depict the accumulated response time starting with the reception of a client request at a controller replica until the last reconfiguration in one of the switches on the detected path. The total number of active controllers was fixed to $|\mathcal{C}|=10$ and the measurement was executed for A\&E group sizes varying between $|\mathcal{A}|=3$ and $|\mathcal{A}|=7$ ($F_M = 1$ and $F_M =3$, respectively) controllers. Rejecting executions were not considered. Both Fat-Tree and Internet2 topologies depict the benefit of opportunistic execution and a lower number of phases in OBFT in all scenarios.

\begin{figure}[htb]
	\centering
	\subfloat[Fat-Tree Topology]{
		\centering
		\includegraphics[width=0.49\textwidth]{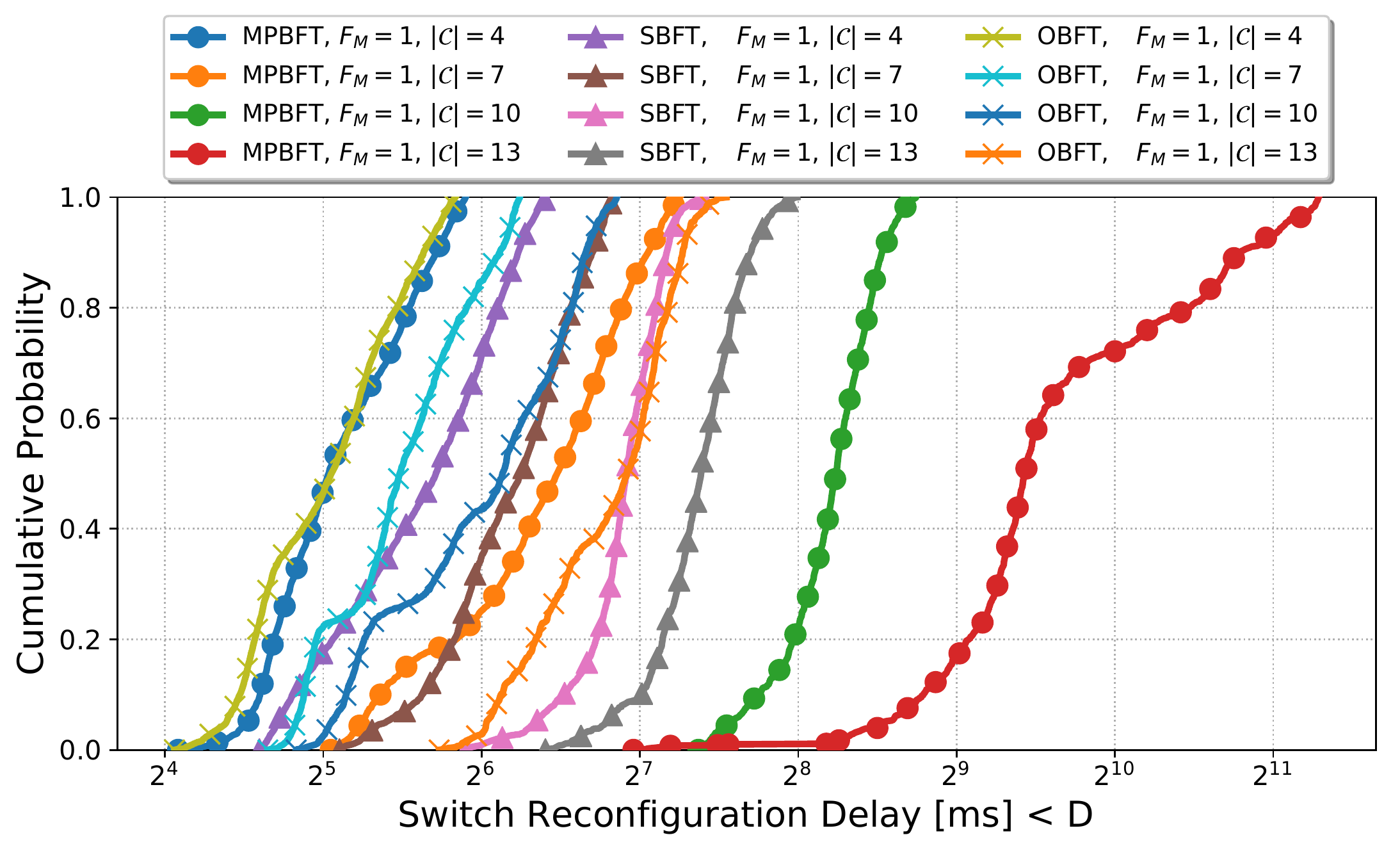}
		\label{fig2:fattree}
	}
	\newline
	\subfloat[Internet2 Topology]{
		\centering
		\includegraphics[width=0.49\textwidth]{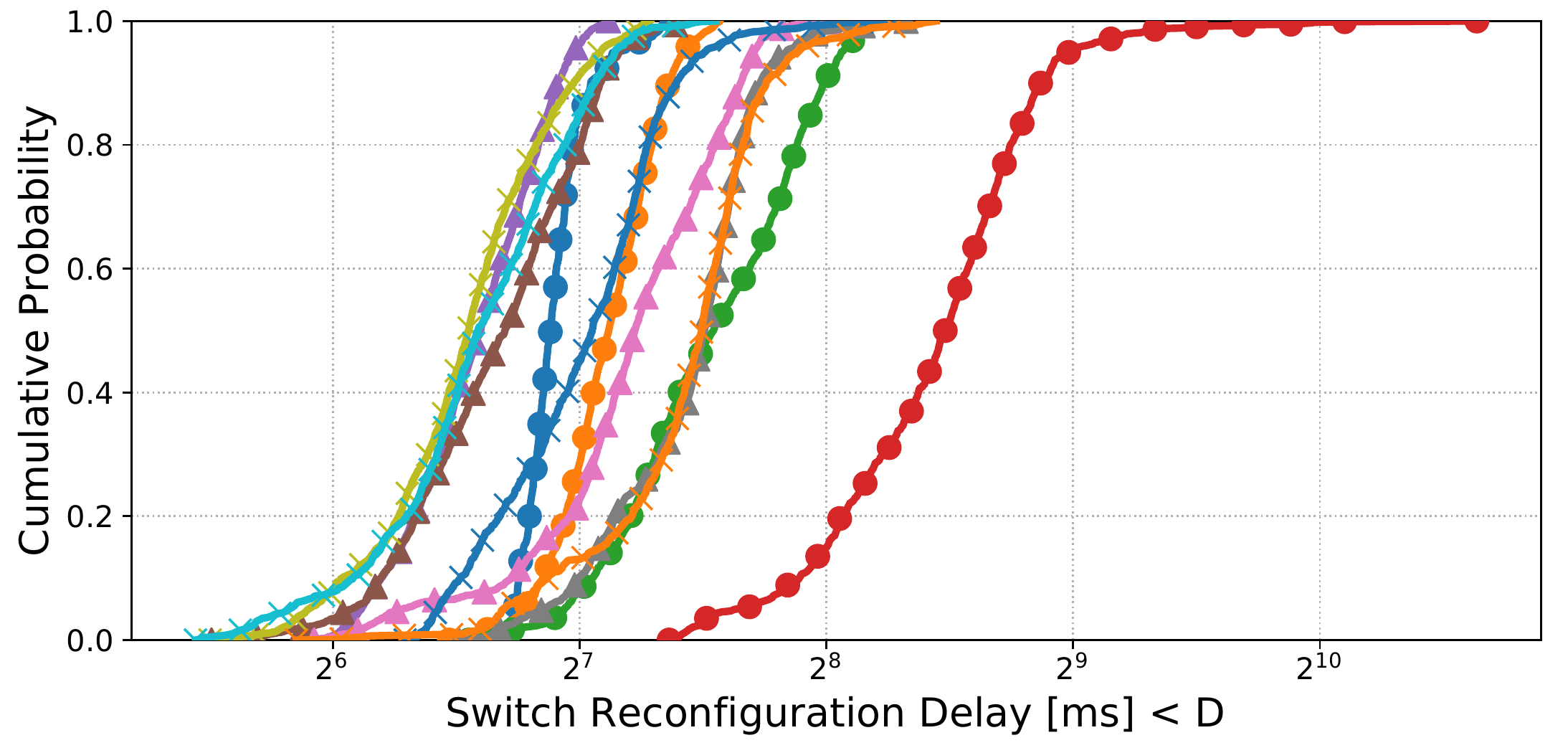}
		\label{fig2:internet2}
	}
	\caption{Total accumulated switch reconfiguration (system response) time for varied numbers of active controller replicas $|\mathcal{C}| = [4..13]$,  $F_A=0$ and an A\&E group size of $|\mathcal{A}|=3$. While OBFT portrays the lowest reconfiguration delays, its performance is similar to SBFT and MPBFT for small control planes (especially for Internet2), slightly better compared to SBFT and largely dominant compared to MPBFT for larger topology sizes.}
	\label{fig:c2sdelaymetrics2}
\end{figure}

In Fig. \ref{fig2:fattree} and Fig. \ref{fig2:internet2}, we vary the total number of deployed active controllers. The figures portray how MPBFT provides equal performance for the controller constellations where the A\&E group size in SBFT and OBFT approximately equals the total number of active controllers (all controllers belong to the same A\&E group). After provisioning additional replicas (case for $|\mathcal{C}| = [7..13]$), the performance of MPBFT starts to suffer compared to both SBFT and OBFT, as it requires interactions between all instances of controllers for successful request handling, whereas SBFT and OBFT continue to operate at the level of a constant A\&E group size. OBFT offers the best performance in both topologies. This is due to SBFT and MPBFT requiring additional rounds to handle the request sequencing, compared to OBFT, that ensures the causality property holds per-switch, even in the case of unordered executions. MPBFT suffers further since the commands execute on each of the controller replicas. Hence, its consensus requires on average an inclusion of a larger number of replicas compared to SBFT and OBFT. Internet2 topology depicts a lower discrepancy between SBFT and OBFT and highlights the benefit of sequencing in geographically distributed scenarios where network delays cause a longer asynchronous period and thus a higher probability of execution overlaps (confirmed by Fig. \ref{fig:acceptancerate}). The maximum path lengths are higher for Internet2 topology, thus resulting in a higher number of overlapping reservations that cause execution rejections/stalling period in opportunistic OBFT.

\subsubsection{Acceptance rates for arriving requests} 
\begin{figure}[htb]
	\centering
	\includegraphics[width=0.49\textwidth]{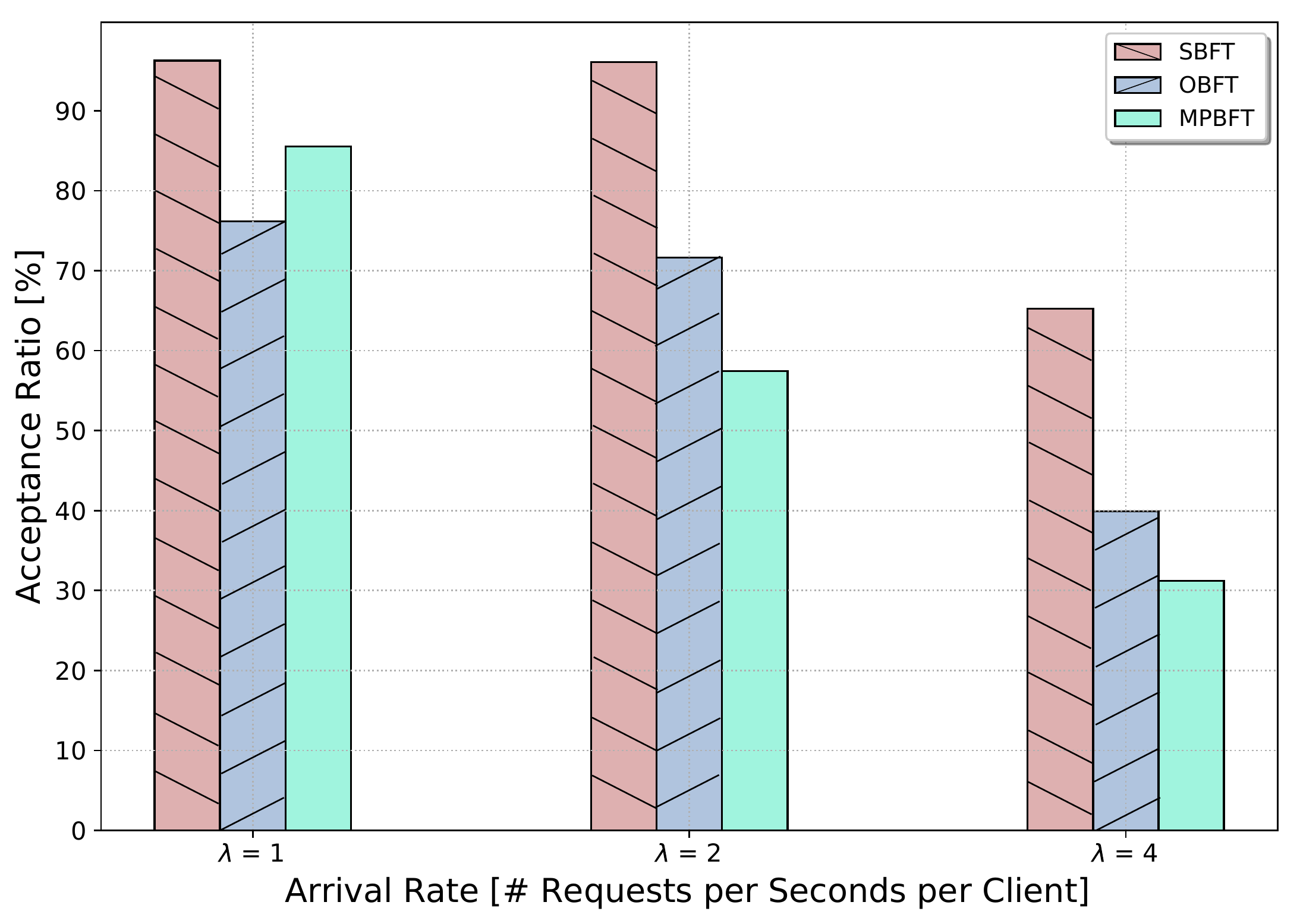}
	\caption{Acceptance rates for incoming client requests in fat-tree topology and $F_M=1, |\mathcal{C}|=4$. SBFT tends to execute a higher number of \emph{successful} runs compared to: i) MPBFT, due to its larger number of active replicas involved in sequencing process and ii) OBFT, due to its opportunistic design, where consistency of outputs is agreed upon after execution has finished.}
	\label{fig:acceptancerate}
\end{figure}

In Fig. \ref{fig:acceptancerate} we vary the per-client arrival rates $\lambda$ for incoming client requests. In the case of $\lambda = 4$, up to 64 requests/second are processed by the cluster in Internet2 topology. Opportunistic execution of OBFT and subsequent hash comparison tends to result more often in rejecting runs, compared to SBFT that serializes all requests prior to their processing. MPBFT results in a relatively high percentage of rejections, due to a higher chance of conflicting sequence number handouts that may occur concurrently since all replicas are involved in proposals during \texttt{PREPARE} phase.

\subsubsection{Communication overhead} 
\begin{figure}[htb]
	\centering
	\includegraphics[width=0.49\textwidth]{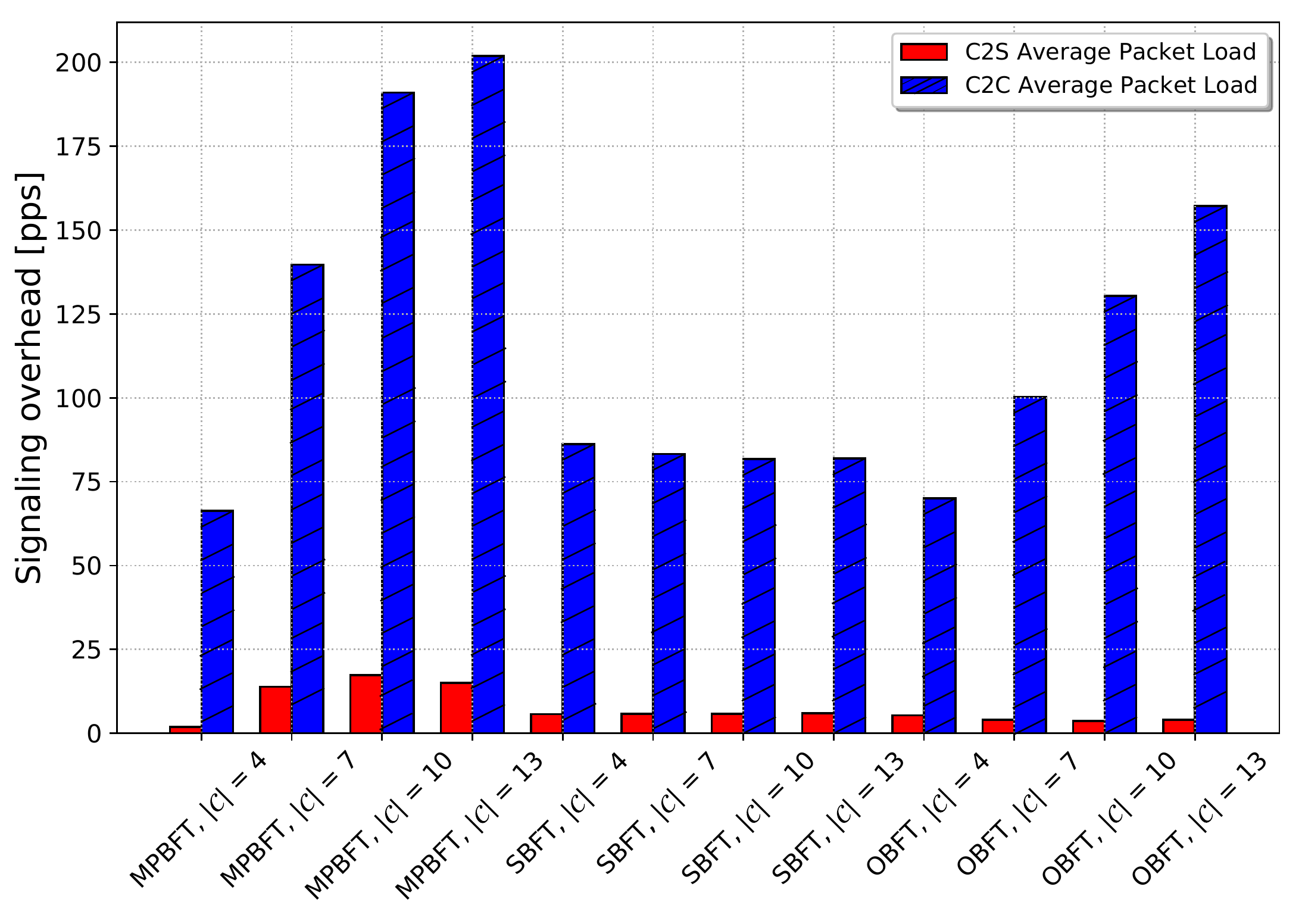}
	\caption{Signaling overhead [pps] when serving 16 requests/second for a varied number of controllers $|\mathcal{C}|=[4..13]$ and a fixed A\&E group size $|\mathcal{A}| = 3$. SBFT possesses the lowest overhead (linear growth), followed by OBFT and MPBFT, that show a quadratic growth scaling with $|\mathcal{C}|$.}
	\label{fig:overhead}
\end{figure}

Fig. \ref{fig:overhead} depicts the scaling of communication overhead with the increase of the total number of active controllers. Controller-to-Switch (C2S) communication overhead increases with the number of controllers that execute the operation and communicate their result to the target switches. Thus, following an output response computation, in MPBFT each controller distributes the newly computed configurations to switches, hence the linear overhead increase. Since the size of the A\&E group remains unchanged throughout all depicted scenarios, SBFT and OBFT show a constant low C2S overhead. The Controller-to-Controller (C2C) overhead scales with the number of active controllers involved in the A\&E group. For MPBFT and OBFT, this increase is quadratic. For SBFT, the C2C overhead increase is linear. It should be noted that the linear evolution holds only for constant A\&E group sizes, i.e., for fixed $F_M$ and $F_A$. 

\emph{Additional notes}: While SBFT and MPBFT ensure a single execution and validation of inputs for client requests (i.e., each client sequence number is mapped to a unique request), OBFT executes client requests speculatively, prior to reaching consensus. Thus, Byzantine clients may attempt affecting the order of execution, or generate execution contentions. Metering mechanisms for misbehaving clients and their exclusion could cater for this case. They are, however, not in the scope of this work.

\section{Related Work}
\label{relatedwork}
Agreement-based approaches have focused on the optimization of sequencing procedure by minimizing the number of replicas that actively participate in sequence proposals \cite{distler2016resource, liu2018scalable}. \emph{REBFT} \cite{distler2016resource} keeps only a subset ($2F+1$ of a total of $3F+1$) replicas active during normal case operation. It activates the passive replicas only after a detected replica fault. Such approaches rely on a trusted counter implementation to prevent \emph{equivocation}, the capability of a malicious replicas to send conflicting proposals to other members. Since we do not assume a centralized proposer, we prevent equivocation by deciding new seq. numbers individually, without the overhead of a trusted counter nor passive replica activation delay.

Speculative BFT protocols have been investigated in \cite{kotla2007zyzzyva, mohan2017primary}. However, these approaches conclude about the consensus of the computed decisions based on the comparison of the instantaneous outputs and assume a \emph{stateless} operation. In the contrast, in OBFT we leverage the agreement procedure that relies on external outputs, i.e., \emph{stateful} per-switch configurations that are inherent to network management scenarios.

Omada \cite{eischer2017} is a sequencing-based BFT design that assigns replicas with either agreement or execution roles and parallelizes the agreement phase. It highlights the benefit of selecting a configuration with the lowest number of agreement groups. Contrary to our work, the authors assume a centralized sequencer per agreement group. Distinguishing causality property per configuration target is not discussed nor leveraged in their protocol. Similarly, Omada does not provide an insight into opportunistic approaches to execution handling.

\section{Conclusion}
\label{conclusion}

We have implemented two agreement-based and an opportunistic BFT protocol for the purpose of SDN controller state synchronization, and have analyzed their overheads in an emulated environment using software switches and emulated network delays. The evaluated KPIs include the switch reconfiguration times, the request acceptance rates and the communication overhead. We have shown how our opportunistic BFT approach leverages agreement of switch state at the time of request computation to ensure the causality during request reconfiguration. It offers considerably lower response time compared to the sequencing-based approaches. However, this benefit comes at the expense of a lower acceptance rate and quadratic communication overhead. For those metrics, the A\&E group-based sequencing approach SBFT presents a better alternative. Both approaches result in a higher throughput compared to MPBFT, which adapts the PBFT protocol.

\section*{Acknowledgment}

This work has received funding from the European Union's Horizon 2020 research and innovation programme under grant agreement number 780315 SEMIOTICS. We are grateful to Nemanja Deric, Arled Papa, Johannes Riedl and the reviewers for their useful feedback and comments.



%



\bibliographystyle{IEEEtran}
\bibliography{IEEEabrv,qos}

\end{document}